\documentclass[11pt]{article}
\usepackage{amsmath}
\usepackage{amsfonts}
\usepackage{amssymb}
\usepackage{epsfig}
\usepackage{times}
\newcommand{\rar}{\rightarrow}
\newcommand{\ol}{\overline}
\newcommand{\ti}{\tilde}
\newcommand{\bs}{\boldsymbol}

\renewcommand{\l}{\newline\null}
\makeatletter
\renewcommand{\section}{\@startsection
{section}%
{1}%
{0em}%
{-1mm}%
{1mm}%
{\Large\bfseries}}%
\renewcommand{\subsection}{\@startsection
{subsection}%
{2}%
{0em}%
{-1mm}%
{1mm}%
{\bfseries}}%
\begin{document}
%
%%%%%%%%%%%%%%%%%%%%%%%%%%%%%%%%%%%%%%%%%%%%%%%%%%%%%%%%%%%%%%%%%%%%%%%%%%%%%
\begin{titlepage}
March 2001 (revised May 2001)\hfill  PAR-LPTHE 01/15\l
\hfill  Ref.~SISSA 28/2001/EP
%\begin{flushright} hep-ph/0103334 \end{flushright}
\vskip 3.5cm
{\baselineskip 17pt
\begin{center}
{\bf HIERARCHIES OF QUARK MASSES AND THE MIXING MATRIX\break
IN THE STANDARD THEORY}
\end{center}
}
\vskip .3cm
\centerline{
B. Machet
     \footnote[1]{LPTHE, tour 16\,/\,1$^{er}\!$ \'etage,
          Universit\'e P. et M. Curie, BP 126, 4 place Jussieu,
          F-75252 Paris Cedex 05 (France).\l
\hskip 1cm{\em Unit\'e associ\'ee au CNRS  et aux Universit\'es Paris 6
(P. et M.  Curie) et Paris 7 (Denis Diderot): UMR 7589}
}
     \footnote[2]{Email: machet@lpthe.jussieu.fr}
\&\ \ S.T. Petcov
     \footnote[3]{SISSA, Via Beirut 2-4, I-34013 Trieste (Italy)}
     \footnote[4]{E-mail: petcov@he.sissa.it}
     }
\vskip 1.5cm
%
%%%%%%%%%%%%%%%%%%%%%%%%%%%%%%%%%%%%%%%%%%%%%%%%%%%%%%%%%%%%%%%%%%%%%%%%%%%
%
{\bf Abstract.}  We study the general dependence of  mixing angles on
heavy fermion masses when  mass hierarchies exist among the fermions.
For two generations and small Cabibbo angle, this angle is directly shown to
scale like $\mu_1/m_s \pm \mu_2/m_c$, where
$\vert\mu_1\vert \ll m_s, \vert\mu_2\vert \ll m_c$ are independent
mass scales.
For $n=3$ generations, we extend to the Yukawa matrices of $u$- and $d$-type
quarks the property that the $2\times 2$
upper-left sub-matrix of the Cabibbo-Kobayashi-Maskawa  matrix $K$ is a good
approximation to the Cabibbo matrix $C$. Then, without any additional Ansatz
concerning the existence of mass hierarchies or the smallness of
the mixing angles, the moduli of its entries $K_{13},K_{23},K_{31},K_{32}$ are
shown to scale like
$[\beta_{13},\beta_{23},\beta_{31},\beta_{32}] \sqrt{{m_c}/{m_t}} \pm
[\delta_{13},\delta_{23},\delta_{31},\delta_{32}] \sqrt{{m_s}/{m_b}}$,
where the $\beta$'s and the $\delta$'s are coefficients
smaller than $10$.
This method, when used for two generations, gives a dependence
on $m_s$ and $m_c$ ``weaker'' 
than the one obtained first, but which matches a well known behaviour
for the Cabibbo angle: $\theta_c \approx
\sqrt{\epsilon_d (m_d/m_s)} - \sqrt{\epsilon_u(m_u/m_c)}$, with
$\epsilon_d,\epsilon_u \leq 1$.
The asymptotic behaviour in the case of three generations can also
be strengthened into a $1/m_{b,t}$ behaviour by incorporating our knowledge
about the hierarchies of quark masses and the smallness of the mixing angles.

%%%%%%%%%%%%%%%%%%%%%%%%%%%%%%%%%%%%%%%%%%%%%%%%%%%%%%%%%%%%%%%%%%%%%%%%%%%

\smallskip

{\bf PACS:} 12.15.Ff
\vfill
\vskip 2.5cm
\centerline{
\epsfig{file=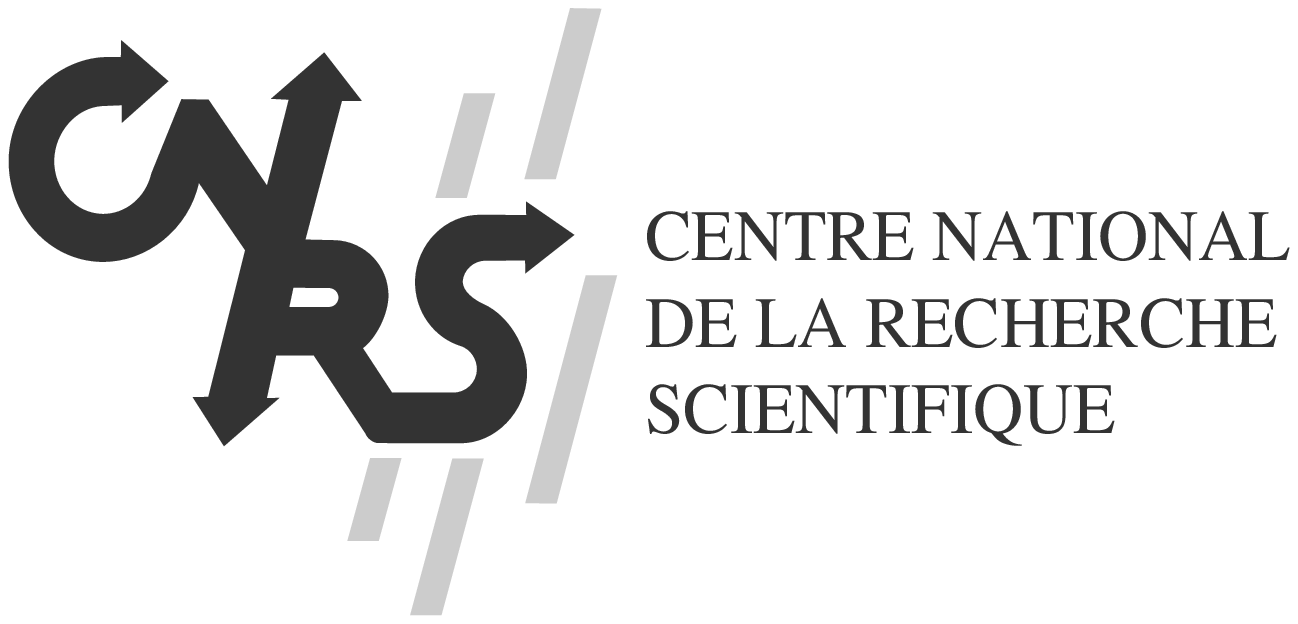,height=2truecm,width=4.5truecm}
\hskip 1cm
\epsfig{file=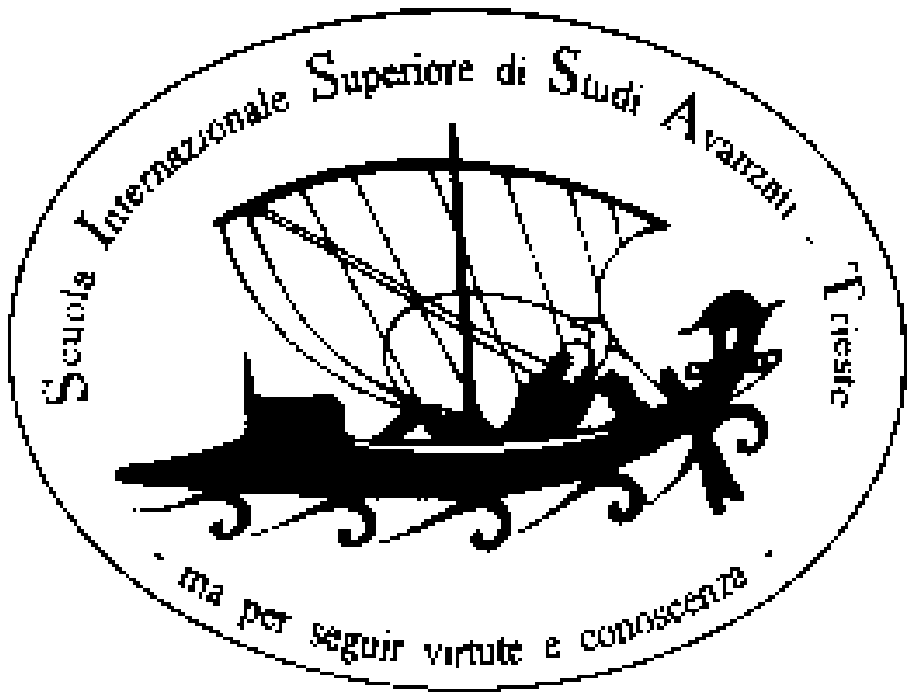,height=2.3truecm,width=3.5truecm}
}
\end{titlepage}
%
%%%%%%%%%%%%%%%%%%%%%%%%%%%%%%%%%%%%%%%%%%%%%%%%%%%%%%%%%%%%%%%%%%%%%%%%%%%%%
%
%%%%%%%%%%%%%%%%%%%%%%%%%%%%%%%%%%%%%%%%%%%%%%%%%%%%%%%%%%%%%%%%%%%%%%%%%%%%
\section{Introduction}
\label{section:introduction}
%%%%%%%%%%%%%%%%%%%%%%%%%%%%%%%%%%%%%%%%%%%%%%%%%%%%%%%%%%%%%%%%%%%%%%%%%%%%
The fermionic mass spectrum and the mixing angles are intimately
connected in the Standard Model \cite{GlashowSalamWeinberg}.
However, the number of independent (Yukawa) couplings being larger than
the total
number of masses and mixing angles, one cannot deduce, without additional
assumptions concerning flavour physics beyond the Standard Model,
one-to-one relationships between mixing angles and (ratios of)
fermion masses.
In general, any relation between them involves
arbitrary entries of the Yukawa matrices; renormalization group arguments
\cite{Pokorski}\cite{RamondRobertsRoss}\cite{PecceiWang}
add to this indetermination  by showing that  the latter
can undergo large variations when the largest mass scale runs
from the electroweak scale to the unification scale.
The presence of a number of zeroes in the Yukawa matrices at the
unification scale can eventually reduce the number of independent
variables to the one of measurable quantities, and this possibility
was extensively studied 
\cite{RamondRobertsRoss}\cite{GeorgiJarlskog}\cite{Fritzsch}\cite{FritzschXing}
\cite{FalconeTramontano}
\cite{PecceiWang} \cite{KuoWu}\cite{Desai}\cite{Branco}\cite{Berger}.
The missing information could also be supplied by introducing
additional flavour symmetries into the Standard Model or its extensions
(see for example \cite{BarbieriHallRomanino} and references therein).

In the present article,  we derive  general ``asymptotic''
relations between mixing angles and quark masses, valid, 
strictly speaking,
when the two heaviest quarks of the $u$- and $d$- type have 
masses much larger
than the others. These relations are obtained  from
the following requirements, the first being dictated by experiment:\l
- there exist hierarchies between the quark masses inside each family;\l
- one must be able to describe correctly, {\em i.e.} accurately and
coherently, physics with a given number of observed families, even if
there exist additional families.

In the case of two generations, taking into account the existence of
separate hierarchies between the masses of the two charge $2/3$ quarks
and the masses of the two charge $-1/3$ quarks,
and the fact that the Cabibbo angle is small, we show that the latter
behaves like $\theta_c \approx \mu_1/m_s \pm \mu_2/m_c$ when $m_s \gg m_d$
and $m_c \gg m_u$, where $m_s, m_d$, $m_c$, $m_u$ are respectively
the masses of the $s$, $d$, $c$ and $u$ quarks, $\mu_{1,2}$ are real
independent
mass scales and $\vert\mu_1\vert \ll m_s, \vert\mu_2\vert\ll m_c$.

As the method used for $n=2$ generations cannot be straightforwardly extended
to a larger number of families, for $n=3$, we do not make any hypothesis
concerning the existence of hierarchies or the smallness of the
mixing angles but, instead, we
introduce another requirement,  shown to be very mild:
like the $2\times 2$ upper-left sub-matrix of
the Cabibbo-Kobayashi-Maskawa (CKM) matrix $K$ \cite{CKM}
matches the Cabibbo matrix to a very good approximation, we require that,
for each ($u$ and $d$) type of quarks, taking as Yukawa mass matrix for two
generations the upper-left
$2\times 2$ sub-matrix of the one for three generations yields
an uncertainty in the mass spectrum (of two generations) which is much
smaller than the larger of the two quark masses.
This condition yields by itself interesting  constraints on
the asymptotic behaviour of mixing angles.
For two generations, one recovers the well-known result that the
Cabibbo angle behaves, when $m_s \gg m_d, m_c \gg m_u$ like 
$\sqrt{\epsilon_d (m_d/m_s)} - \sqrt{\epsilon_u(m_u/m_c)}$,
where $\epsilon_d, \epsilon_u \leq 1$.
For $n=3$, a similar behaviour is demonstrated 
for the elements $K_{13}, K_{31}, K_{23}, K_{32}$ of the CKM mixing matrix.

The above ``stability'' of the mixing matrix as one goes from $n$ to $(n \pm
1)$ generations could be fortuitous:
nature may be such that the whole structure of mass matrices for $(n-1)$
generations is totally spoiled when one goes to $n$
\footnote{As this is evidently untrue for their diagonal form, this
 would tend to emphasize the role of non-diagonal elements.}
; one then would probably face the fact that, if there exist generations
beyond the ones presently known, the
knowledge about Yukawa matrices  established at our
energies is mostly illusory and, from it, we cannot expect 
reliable hints for the ``complete'' theory. In particular, this would spoil
the renormalization group arguments which evolve Yukawa matrices to the
unification scale \cite{Pokorski}\cite{RamondRobertsRoss}\cite{PecceiWang},
where one can expect higher generations.
As  can be easily seen, ``stable'' Yukawa mass matrices yield
``stable'' mixing matrices; we go further and demand here that
 the ``stability'' of the Yukawa matrices be a necessary condition for
the one of the mixing matrix.

Adding the information  that there exist hierarchies for the
masses of the $u$- and $d$- type quarks and that the mixing angles are small,
as represented in a Wolfenstein-like parameterization \cite{Wolfenstein}
 of the CKM mixing matrix, this behaviour can be strengthened into a 
$1/m_{b,t}$ behaviour, like for two generations.
%
%%%%%%%%%%%%%%%%%%%%%%%%%%%%%%%%%%%%%%%%%%%%%%%%%%%%%%%%%%%%%%%%%%%%%%%%%%%
\section{Basic properties of Yukawa matrices in the Standard Model}
\label{section:basic}
%%%%%%%%%%%%%%%%%%%%%%%%%%%%%%%%%%%%%%%%%%%%%%%%%%%%%%%%%%%%%%%%%%%%%%%%%%%
%
The following well known properties  concerning the Standard Model will be
extensively used in the paper.
Let $M_u$ and $M_d$ be the Yukawa mass matrices for the $u$- and $d$-type
quarks.

$\bullet$\ {\em(i)}\  {\em A priori}, $M_u$ and $M_d$ have no special property
of symmetry, and their eigenvalues (defined as the roots of their
characteristic equations) are not the quark masses; the latter are determined
by diagonalizing $MM^\dagger$ and $M^\dagger M$, which are hermitian,
by unitary $U$ and $V$ matrices, according to:
\begin{equation}
U_u^\dagger M_uM_u^\dagger U_u = D_u^2,\quad
V_u^\dagger M_u^\dagger M_u V_u = D_u^2, \quad
U_d^\dagger  M_dM_d^\dagger  U_d = D_d^2,\quad
V_d^\dagger  M_d^\dagger M_d V_d = D_d^2.
\label{eq:UV}
\end{equation}
The entries of the diagonal matrices $D_u$ and $D_d$ are the quark
masses. Their signs  are irrelevant. $M_u$ and $M_d$ are
brought to their diagonal forms by bi-unitary transformations
\begin{equation}
U_u^\dagger  M_u V_u = D_u,\quad U_d^\dagger  M_d V_d = D_d,
\label{eq:diagM}
\end{equation}
corresponding to acting on left-handed fermions with the $U$'s and on
right-handed ones with the $V$'s.
The CKM mixing matrix is then
\footnote{ A trivial consequence of (\ref{eq:CKM}) is that, in
a basis where $M_u$ is diagonal ($U_u = {\mathbb I} = V_u$), $K = U_d$.}
\begin{equation}
K = U_u^\dagger  U_d.
\label{eq:CKM}
\end{equation}
The unitarity of $U$ and $V$ ensures that the kinetic terms keep diagonal.

$\bullet$\ {\em(ii)}\  Arbitrary independent rotations can be performed on
$u$- and $d$-type right-handed quarks ({\it i.e.} quark fields),
and identical transformations on left-handed $u$- and $d$-type
quarks.

$\bullet$\ {\em(iii)}\  One can swap columns in $M$ by a rotation
on right-handed quarks (equivalent to right-multiplying $M \rightarrow MR$);
one can swap lines  by a rotation on left-handed quarks, but
the same rotation must be done on $u$- and $d$-type quarks
($M_u \rightarrow LM_u, M_d \rightarrow LM_d$).

$\bullet$\ {\em(iv)}\  Any mass matrix can be brought to a triangular form by
rotations on right-handed quarks.

$\bullet$\ {\em(v)}\  {\em Polar decomposition
theorem}\cite{TurnbullAitken} \cite{FramptonJarlskog}:
 any complex  matrix $M$ can be written like
the product of an hermitian  matrix $(MM^\dagger)^{1/2}$ times a
unitary  matrix; in the Standard Model, the latter can always
be absorbed  by a rotation on right-handed quarks; hence, there,
all mass matrices can always be taken to be hermitian;
one has then $U=V$, and the quark masses become identical to the eigenvalues of
the mass matrices.

$\bullet$\ {\em(vi)}\  $K$ (\ref{eq:CKM}) is left invariant if $U_u$ and $U_d$
are multiplied by the same unitary matrix $A$; this is also true for the
corresponding right-handed mixing matrix $K_R = V_u^\dagger V_d$
(of no relevance in the Standard Model) when $V_u$ and $V_d$ are multiplied
by $A$; thus $M_u$ and $M_d$ are determined up to a common unitary
transformation
$M_u \rightarrow A^\dagger M_u A, M_d \rightarrow A^\dagger M_d A$.
%
%%%%%%%%%%%%%%%%%%%%%%%%%%%%%%%%%%%%%%%%%%%%%%%%%%%%%%%%%%%%%%%%%%%%%%%%%%%
\section{The case of two generations}
\label{section:2gen}
%%%%%%%%%%%%%%%%%%%%%%%%%%%%%%%%%%%%%%%%%%%%%%%%%%%%%%%%%%%%%%%%%%%%%%%%%%%
%
We investigate here the relation between the Cabibbo angle $\theta_c$ and
the quark masses.
The case of two generations is considered separately and can be treated
in a simple way which only supposes the existence of hierarchies among each
type of quarks and uses the observed smallness of the mixing angle.
The Yukawa matrices can be taken to be real, which we assume here.

First, we exploit the existence of  mass hierarchies for the
$u$- and $d$-type quarks, {\it i.e.} that $m_s \gg m_d$ and $m_c
\gg m_u$.
For each type of quarks, this hierarchy of masses requires that
at least one of the entries in the
corresponding Yukawa matrix $M$ must be larger than -- or equal to -- the
larger of the two quark masses;
otherwise all eigenvalues of $MM^T$ and $M^T M$
\footnote{The superscript ``$T$'' means ``transposed''.}
 would be smaller
than the scale $\Lambda$ set by the larger quark mass.
$M$ can have one, two, three or four entries larger or equal to $\Lambda$.
The case of four large entries can always be reduced to three by a right-handed
rotation which, for example, brings $M$ to the upper triangular form.
If there are three large entries, the sum and the product of the eigenvalues of
$MM^T$ are large, which mean that both quark masses are larger than
$\Lambda$.

The case with two large entries can be further reduced in that they
cannot be both on the diagonal, since, then, the two quark masses are again
larger than $\Lambda$. Furthermore, they  must be on the
same column. Suppose indeed that they are on the same line. A right-handed
rotation can then always reduce the number of large entries to one.
For instance
\footnote{Throughout the paper the use the abbreviated notations $c$ for
$\cos$ and $s$ for $\sin$. The subscripts referring to the corresponding
angles should avoid any confusion with the the charm and strange quarks.}
\begin{equation}
MR = \left(\begin{array}{cc} \Lambda_1 & \Lambda_2 \cr
                                 a     &     b \end{array}\right)
          \left(\begin{array}{rr} c_\varphi  &  s_\varphi  \cr
                                  -s_\varphi  &  c_\varphi \end{array}\right)
\rightarrow c_\varphi\left(\begin{array}{cc}
       \Lambda_1 + \Lambda_2^2/\Lambda_1 &  0 \cr
       a + b \Lambda_2/\Lambda_1 & -a \Lambda_2/\Lambda_1 +b
\end{array}\right)  \quad \text{for}\ s_\varphi = -c_\varphi \Lambda_2/\Lambda_1
\label{eq:MR}
\end{equation}
where $\vert\Lambda_{1,2}\vert \geq \Lambda$ and $\vert a \vert, \vert b
\vert \ll \Lambda$, has only one large entry at the $[1,1]$ position.
The same argument applies  when the two initial
large entries are on the second line.

Suppose, accordingly,  that the two large entries are on the same column,
and  use the freedom to make common left-handed rotations on both $u$
and $d$ type quarks:
\begin{equation}
LM = \left(\begin{array}{rr} c_\varphi  &  s_\varphi  \cr
                                  -s_\varphi  &  c_\varphi \end{array}\right)
   \left(\begin{array}{cc} \Lambda_1 & a \cr
                           \Lambda_2 & b \end{array}\right)
\rightarrow c_\varphi \left(\begin{array}{cc}
  \Lambda_1 + \Lambda_2^2/\Lambda_1 & a + b\Lambda_2/\Lambda_1 \cr
                        0        &  -a\Lambda_2/\Lambda_1 + b
\end{array}\right)  \quad \text{for}\ s_\varphi = c_\varphi \Lambda_2/\Lambda_1;
\label{eq:LM}
\end{equation}
$LM$ has now only one large entry at the $[1,1]$ position and this rotation
does not alter the structure of the second mass matrix,  supposed to
also have two large entries on the same column (whatever be the column).
So we can state:

{\em For two generations, 
the existence, for each type of quarks,
of one mass larger than a certain scale, the other being much smaller
than this scale,
entails that one can restrict to Yukawa matrices having either one large
entry each, or to the case where one of them  has two large entries in the same
column and the other has only one large entry; by a right-handed rotation,
the  latter can be put in any column.}

$\bullet$\quad Accordingly, we consider, first, for example,
$M_u$ with two large entries on the same column and $M_d$ with only one
\begin{equation}
M_u = \left(\begin{array}{cc} \Lambda_1  &  a \cr
                              \Lambda_2  &  b \end{array}\right),\quad
\vert\Lambda_1\vert,\vert\Lambda_2\vert \gg \vert a\vert,\vert b\vert,\quad
M_d = \left(\begin{array}{cc}    u       &   v   \cr
                                 w       &   \Lambda_3 \end{array}\right),
\quad \vert\Lambda_3 \vert\gg \vert u\vert,\vert v\vert,\vert w\vert.
\label{eq:MuMd}
\end{equation}
The successive  steps are the following. %:\ - 
From (\ref{eq:CKM}), the Cabibbo angle can be written as
\begin{equation}
\theta_c = \theta_d - \theta_u. % ;
\label{eq:thetac}
\end{equation}
Using (6) and diagonalizing $M_d M_d^T$ we get
\begin{equation}
\tan (2\theta_d) = 2 \frac {uw + v\Lambda_3}{\Lambda_3^2 + w^2 -u^2-v^2}. %;
\label{eq:tand}
\end{equation}
Thus, when $\vert\Lambda_3 \vert\gg \vert u\vert,\vert v\vert,\vert w\vert$,
$\theta_d$ is small and behaves like
 $ \theta_d \approx v/\Lambda_3$. % ;\l
From (6) one finds similarly
\begin{equation}
\tan (2\theta_u) = 2 \frac {ab + \Lambda_1\Lambda_2}
{\Lambda_2^2 - \Lambda_1^2 + b^2 -a^2}. % ;
\end{equation}
We now use as input our knowledge about the Cabibbo angle: 
as $\theta_c$ is experimentally small and $\theta_d$ has just been proven
  to be small, (\ref{eq:thetac}) requires that $\theta_u$ is also small.
This can only occur if  $\vert\Lambda_1 \vert\gg \vert\Lambda_2\vert$ or
$\vert\Lambda_2 \vert\gg
\vert\Lambda_1\vert$.
Let us consider, for example, $\vert\Lambda_1\vert \gg
\vert\Lambda_2\vert$. % ;\l
Then $\theta_u$ behaves like $-\Lambda_2/\Lambda_1$. % ;\l
Thus, $\theta_c$ behaves like $v/\Lambda_3 + \Lambda_2/\Lambda_1$. % ;\l
The quark masses are
\begin{equation}
\vert m_d \vert \approx \vert u \vert,\quad \vert m_s \vert \approx \vert
\Lambda_3 \vert,\quad
\vert m_u \vert \approx \frac{\vert a\Lambda_2 -
b\Lambda_1\vert}{\sqrt{\Lambda_1^2 + \Lambda_2^2}} \approx \vert b
\vert,\quad
\vert m_c \vert \approx \sqrt{\Lambda_1^2 + \Lambda_2^2} \approx
\vert\Lambda_1 \vert.
\label{eq:qmass}
\end{equation}
The limits $m_s \gg m_d, m_c \gg m_u$ do not depend on the non-diagonal
elements $a$, $\Lambda_2$ $v$, $w$ of the mass matrices, which
stand, in this limit, as  independent entries only determined
by physics of flavour beyond the Standard Model.
This allows the claim: given the fact that $m_s \gg m_d,  m_c \gg  m_u$,
\begin{equation}
\theta_c \approx \frac{v}{m_s} \pm \frac{\Lambda_2}{m_c},\quad m_s \gg \vert v
\vert, m_c \gg \vert \Lambda_2 \vert.
\label{eq:limthetac}
\end{equation}
$\bullet$\quad The case when the two mass matrices $M_u$ and $M_d$
have only one large
entry each is trivial since, then, one also gets a small $\theta_u$ and the
equivalent of (\ref{eq:tand}) for $\tan(2\theta_u)$.

$\bullet$\quad One can consequently state the general result for the Cabibbo angle:
\begin{equation}
\theta_c \approx \frac{\mu_1}{m_s} \pm \frac{\mu_2}{m_c},\quad 
         m_s \gg \vert\mu_1\vert, m_c \gg \vert \mu_2 \vert,
\end{equation}
where $\mu_{1,2}$ are independent real  mass scales 
and we have neglected
corrections in higher inverse powers of $m_s$ and $m_c$.
The mass scales $\mu_{1,2}$ depend on the basis.
If one goes to a basis where, e.g.,
$M_u$ is diagonal, then $\theta_u = 0$ and $\mu_2 = 0$. This reflects the 
similar roles held by the two members inside  an $SU(2)$ doublet and, in
particular, that one cannot push one of the two masses alone to infinity
without breaking the renormalizability of the theory
\footnote{One has then to implement  a non-linear
realization of the gauge symmetry \cite{FeruglioMasieroMaiani}}
. 
%%%%%%%%%%%%%%%%%%%%%%%%%%%%%%%%%%%%%%%%%%%%%%%%%%%%%%%%%%%%%%%%%%%%%%%%%%%
\section{Defining hierarchical matrices; algebraic properties}
\label{section:defhier}
%%%%%%%%%%%%%%%%%%%%%%%%%%%%%%%%%%%%%%%%%%%%%%%%%%%%%%%%%%%%%%%%%%%%%%%%%%%
Consider the following set $\{{\cal M}\}_h$ of real $n \times n$ matrices
$\cal M$ ($n$ is here the number of generations of fermions) such that,
for any of them:\l
- the modulus of its diagonal element ${\cal M}_{nn}$ is of order unity
   $\vert{\cal M}_{nn}\vert \approx 1$:\l
- its non-diagonal border ( {\it i.e.} in the lowest line and extreme
  right column) elements are small in the following sense:
${\cal M}_{in} = \lambda_i/\Lambda_i, {\cal M}_{nj} = \xi_j/\Xi_j,
i,j = 1\cdots (n-1),
\vert\lambda_i\vert \ll \vert\Lambda_i\vert, \vert\xi_j\vert \ll
\vert\Xi_j\vert$;\l
- its other elements ${\cal M}_{ij},i,j = 1 \cdots (n-1)$ satisfy
\hbox{$\vert {\cal M}_{ij}\vert _{i,j = 1 \cdots (n-1)} \leq 1$}.
 
We shall call hereafter such matrices {\em hierarchical}
\footnote{This definition  constrains only the
non-diagonal elements on the border to be small.}
.
 
{\bf Property 1}:  Neglecting terms of second order in
$\lambda_i/\Lambda_i$ and $\xi_j/\Xi_j$, and considering the usual
multiplication of matrices:\l
- the product of two hierarchical matrices is hierarchical;\l
- the multiplication of hierarchical matrices is of course associative;\l
- the unit matrix, which is hierarchical, is the identity element;\l
- the inverse of a hierarchical matrix is also hierarchical.
 
$\{{\cal M}\}_h$ would form a group, but for the stability
which is not ensured for the product of a very large number of hierarchical
matrices.
 
The CKM mixing matrix $K$ being given by (\ref{eq:CKM})
the properties just stressed entail that if ${U}_u$ and
${U}_d$ are hierarchical, then so is $K$.
 
{\bf Property 2}:
{\it in the same limit of neglecting terms of second order
in $\lambda_i/\Lambda_i$
and $\xi_j/\Xi_j$, the $(n-1)\times (n-1)$ sub-matrix $\tilde{\cal M}$ with
entries $\tilde{\cal M}_{ij} = {\cal M}_{ij},
i=1\cdots (n-1), j=1 \cdots (n-1)$ of a hierarchical
$n\times n$ matrix $\cal M$ which is the product of two $n \times n$
hierarchical matrices $\cal U$ and $\cal V$ only depends of the two
corresponding $(n-1) \times (n-1)$ sub-matrices $\tilde{\cal U}$ and
$\tilde{\cal V}$ of $\cal U$ and $\cal V$}.
 
A consequence is that
the hierarchical structure of the Cabibbo matrix $C$ can be studied
without reference to the third generation after the CKM matrix has been proven
to be itself hierarchical;
explicitly, $K$ is first proven to be hierarchical
from the property (``upper'' hierarchy) $m_b \gg m_s, m_d, m_t \gg m_c,
m_u$ by proving that ${U}_u$ and ${U}_d$ (suitably normalized)
 are both hierarchical; then $C$ can be proven to
be hierarchical from the property (``lower'' hierarchy)
$m_s \gg m_d, m_c \gg m_u$, without worrying anymore about the third
generation.                     

{\bf Property 3}: {\it any unitary matrix $U$ can be made hierarchical by
left- or right- multiplication}.
Indeed, for example, the equation $UA = U_h$ with $U_h$ hierarchical,
has always a solution $A = U^\dagger U_h$.\l
A consequence  is that:

{\bf Property 4}: {\it in the Standard Model and when the mixing matrix
$K$ is hierarchical, one
can always go to a basis where $U_u, U_d$ and the two (hermitian)
mass matrices $M_u, M_d$, normalized to the largest scale available (see
below), are hierarchical}.

Let us first go in a basis where $M_u$ and $M_d$ are hermitian, which is
always possible by the polar decomposition theorem (property {\em (v)} of
section \ref{section:basic}).
Then, we multiply each of the two unitary matrices $U_u$ and $U_d$ by the
same unitary matrix $B$, which, as stressed in section \ref{section:basic}
(property {\em(vi)}), leaves the mixing matrix $K$ invariant.
One chooses $B$ such that $BU_u = U_{uh}$ is hierarchical, which is always
possible as mentioned above;  $BU_d = U_{dh}$ is then also hierarchical
since $U_{dh} = KU_{uh}$, $K$ has been supposed to be hierarchical, and
hierarchical matrices are stable by multiplication.

These transformations on $U_u$ and $U_d$ are equivalent to a change of basis
for quarks $M_u \rar BM_u B^\dagger $, $M_d \rar BM_d B^\dagger $ since the
diagonalisation equations for $M_u$ and $M_d$ write
$D_u = U_u^\dagger  M_u U_u = U_{uh}^\dagger  (BM_u B^\dagger ) U_{uh}$ and
$D_d = U_d^\dagger  M_d U_d = U_{dh}^\dagger  (BM_d B^\dagger ) U_{dh}$;
let us go to the mass matrices normalized to the largest entry
$m_{un}$ of $D_u$ and $m_{dn}$ of $D_d$:
${\cal D}_u = D_u/m_{un}$,
${\cal M}_{uh}= U_{uh}^\dagger  (BM_u B^\dagger ) U_{uh}/m_{un}$,
${\cal D}_d = D_d/m_{dn}$,
${\cal M}_{dh}=U_{dh}^\dagger  (BM_d B^\dagger ) U_{dh}/m_{dn}$;
from ${\cal D}_u = U_{uh}^\dagger  {\cal M}_{uh} U_{uh}$,
${\cal D}_d = U_{dh}^\dagger  {\cal M}_{dh} U_{dh}$, from the fact that
${\cal D}_u$ and ${\cal D}_d$ are hierarchical, and from the stability
property of hierarchical matrices, one deduces that ${\cal M}_{uh}$ and
${\cal M}_{dh}$ are hierarchical.
%
%%%%%%%%%%%%%%%%%%%%%%%%%%%%%%%%%%%%%%%%%%%%%%%%%%%%%%%%%%%%%%%%%%%%%%%%%%%
\section{The case of $\bs{n \geq 2}$ generations}
\label{section:3gen}
%%%%%%%%%%%%%%%%%%%%%%%%%%%%%%%%%%%%%%%%%%%%%%%%%%%%%%%%%%%%%%%%%%%%%%%%%%%
%
A direct demonstration like the one of section \ref{section:2gen}, 
which uses the
smallness of the mixing angles,
needs, to be extended to $n$ generations, inspecting all possible
configurations of $n\times n$
mass matrices which lead to  mass hierarchies between fermions. If feasible, it
would become extremely tedious and inelegant. This is why we perform the
generalization to $n=3$ generations along another line which does not
require any hypothesis concerning the properties of the mixing matrix. It 
provides, in the $n=2$ case, a well-known constraint 
on $\theta_c$, though weaker than the one deduced in section
\ref{section:2gen}.
%
%%%%%%%%%%%%%%%%%%%%%%%%%%%%%%%%%%%%%%%%%%%%%%%%%%%%%%%%%%%%%%%%%%%%%%%%%%%
\subsection{General requirement for Yukawa matrices}
\label{subsection:general}
In the following, like in section \ref{section:defhier}, $\tilde M$ stands
for the $(n-1) \times (n-1)$
upper left sub-block of any given $n \times n$ matrix $M$.
One takes $M$ hermitian (see section \ref{section:basic});
$\tilde M$ is then also hermitian.
A single unitary matrix $U$ is needed to diagonalize $M$ into $D$.

If one steps up or down one generation from $n$ to $(n+1)$ or $(n-1)$,
one must be able to
find a renormalizable description of the corresponding physics: for
example, before the
third generation was suspected, and discovered, the Cabibbo description of two
generations was perfectly coherent, and our present description of physics
with three generations,  also coherent, is supposed to be very little
influenced by the eventual presence of higher generations.

Keeping the previous notations for $n$ generations, let us step down to
$(n-1)$ generations. According to our requirement, there exist, in this
case, too,  Yukawa matrices $\underline M$,
which can also be taken to be hermitian, and the unitary matrices which
diagonalize them into $\underline D$ are called $\underline U$.

In general, $\underline M \not = \tilde M, \underline U \not = \tilde U$.
Experimentally, however,  the (unitary) Cabibbo matrix
$C \equiv \underline{U^T_u} \underline{U_d}$ is very close to the upper $2
\times 2$ (non-unitary) sub-matrix $\tilde K = \widetilde{U^T_u U_d}$ of the
CKM matrix $K$.

As proposed, we demand that this property of
``stability'' of the mixing matrix be the image of an equivalent property for
the Yukawa matrices:
\begin{equation}
\underline M = \tilde M - \epsilon\; m_{n-1} {\mathbb I}_{n-1},
\label{eq:COND}
\end{equation}
where:\l
- \quad $m_n$ and $m_{n-1}$ are the largest quark masses
of the type considered ($u$ or $d$), respectively for $n$ and
$(n-1)$ generations;
for instance $m_n = m_b\ or\ m_t$ and $m_{n-1} = m_s\ or\ m_c$ for $n=3$;
$m_n = m_s\ or \ m_c$ and $m_{n-1} = m_d\ or\ m_u$ for $n=2$;\l
-\quad $\epsilon$, which can {\em a priori} depend on the entries of
$\tilde M$, is  small in the sense $\epsilon \vert \leq 1$;\l
- \quad ${\mathbb I}_{n-1}$ is any $(n-1) \times (n-1)$ matrix with entries
of moduli not larger than one, for example the $(n-1) \times (n-1)$
identity matrix
\footnote{We shall make in the rest of the paper the choice of the
identity matrix for simplicity, but this choice is not mandatory.}
.

Expanding
\begin{equation}
\tilde M \equiv \widetilde{UDU^\dagger } = \tilde U \underline D \tilde
U^\dagger  + m_n\tilde A,
\label{eq:expand}
\end{equation}
$\tilde A$ is an $(n-1) \times (n-1)$ matrix which contains the elements
of the mixing matrix for which we are seeking information;
one gets from (\ref{eq:COND}) the condition
\begin{equation}
\underline M \equiv  \underline U \underline D \underline U^\dagger 
= \tilde U \underline D \tilde U^\dagger  + m_n\tilde A -\epsilon\; m_{n-1}
{\mathbb I}_{n-1},\quad \vert\epsilon\vert \leq 1, \quad or
\label{eq:gen1}
\end{equation}
\begin{equation}
\tilde A = \frac{\underline M - \tilde U \underline D \tilde U^\dagger  +
\epsilon\; m_{n-1} {\mathbb I}_{n-1}}{m_n} =
\frac{\underline U \underline D \underline U^\dagger 
- \tilde U \underline D \tilde U^\dagger 
+ \epsilon\; m_{n-1} {\mathbb I}_{n-1}}{m_n},\quad \vert\epsilon\vert \leq 1.
\label{eq:gen2}
\end{equation}
(\ref{eq:gen2}) will provide the scaling behaviour for the elements
of the mixing matrix that we are seeking for (see the precise examples below).
For this purpose, we shall use the property that the highest mass scale
occurring in the numerator of the r.h.s.  of (\ref{eq:gen2}) is $m_{n-1}$
and that all the matrix elements involved have, because of unitarity, moduli 
smaller or equal to $1$. So, the modulus of the numerator is $m_{n-1} \times$
({\em coefficient of modulus}\ $\leq 2(n-1)^2 + 1$)
\footnote{The $2(n-1)^2 +1$ comes from the following: each entry in
$\underline{U_d} \underline{D_d} \underline{U_d}^\dagger$ and in
$\tilde{U_d}  \underline{D_d} \tilde{U_d}^\dagger$ is the sum of
$(n-1)^2$ terms, each of them being the product of
a $(mass)$ and of two matrix elements of modulus smaller or equal to $1$;
${\mathbb I}_2$ contributes the last $1$.}
.
As cancellations are
expected between $\underline U \underline D \underline U^\dagger$ and
$\tilde U \underline D \tilde U^\dagger$, this bound is expected not to
be saturated (see subsection \ref{subsection:gen2gen} below.)

Further comments about the condition (\ref{eq:COND}) are due, to
show that the property 4 of section \ref{section:defhier} makes it a very
mild assumption.
First, if a $3 \times 3$ mass matrix $\cal M$ ($\cal M$ 
is the mass matrix $M$
normalized to the largest corresponding quark mass) 
is hierarchical, the eigenvalues of
$\tilde{\cal M}$ differ very little from the 
quark masses for two generations
(see the demonstration below); so, $\tilde{\cal M}$ 
has the right mass spectrum, which  is a necessary 
condition for $\tilde{\cal M} \approx \underline{\cal M}$;
then, as far as masses are concerned, the spectrum of fermions is the only
property which can be detected in the framework of the Standard Model,
in the sense that no criterion exists {\em a priori}, without additional
assumptions concerning, e.g., flavour symmetry, which can select
one among several Yukawa matrices yielding the same spectrum.

Our assumption can be rephrased as follows: in a basis where $\cal M$ is
hierarchical, among all possible
$\underline{\cal M}$'s which lead to the right quark spectrum for $(n-1)$
generations,  the choice of the $(n-1)\times (n-1)$
upper left sub-matrix $\tilde{\cal M}$ for
$\underline{\cal  M}$ is considered to be the natural one.
By stating that the mass matrix for $n$ generations can
be obtained by ``dressing'' the one for $(n-1)$ generations, we express
in particular our view, emphasized in the introduction, that a coherent
description of the physics for $(n-1)$ generations knows very little
about higher generations.

Last,
according to property 4 of section \ref{section:defhier}, ${\cal M}_u$ and 
${\cal M}_d$ can always be chosen as hierarchical (by a common change of
basis on $u$- and $d$-type quarks) if the mixing matrix $K$ is hierarchical.
Hence, in this case, our results do not depend on the basis chosen
for ${\cal M}_u$ and ${\cal M}_d$.

The proposition above concerning the eigenvalues of $\ti{\cal M}$ is easily
proven. Up to an overall normalization factor, a hierarchical hermitian
$\cal M$ can be written as
\begin{equation}
{\cal M} = \left( \begin{array}{ccc}
                   a_{11} &  a_{12} &  \eta_1 \cr
                   \ol{a_{12}} &  a_{22} &  \eta_2 \cr
         \ol{\eta_1} & \ol{\eta_2}  &   1  \end{array} \right),\quad
 \text{with}\quad \vert\eta_1\vert,\vert\eta_2 \vert\ll 1.
\label{eq:carac}
\end{equation}
Its characteristic equation reads $\Delta(\lambda)=0$ with
\begin{equation}
\Delta(\lambda) = (1-\lambda)\tilde\Delta(\lambda) 
   + \lambda(\vert\eta_1\vert^2 + \vert\eta_2\vert^2)
-a_{11}\vert\eta_2\vert^2 -
a_{22}\vert\eta_1\vert^2  + a_{12}\ol{\eta_1}\eta_2
+ \ol{a_{12}} \eta_1 \ol{\eta_2},
\end{equation}
where $\tilde\Delta(\lambda)$ is the characteristic equation for
$\tilde{\cal M}$; (\ref{eq:carac}) shows that, up to order $\eta^2$,
$\Delta(\lambda)$ vanishes  for $\lambda = 1$ and
for the eigenvalues of $\tilde{\cal M}$;
the roots of $\Delta(\lambda) = 0$ build up the quark spectrum
and thus, in particular, up to ${\cal O}(\eta^2)$,
 $\tilde{\cal M}$ has the right spectrum for $(n-1)$ generations.
%
%%%%%%%%%%%%%%%%%%%%%%%%%%%%%%%%%%%%%%%%%%%%%%%%%%%%%%%%%%%%%%%%%%%%%%%%%%%
\subsection{Example 1: the case of two generations}
\label{subsection:gen2gen}
Let us for example consider $d$-type quarks. $M_d$ is a (real)
symmetric $2\times 2$ matrix 
and we have
\begin{equation}
M_d  = U_d D_d U_d^T \quad\text{with}\quad
D_d = \left(\begin{array}{cc} m_d & 0 \cr
                               0  & m_s
\end{array}\right)\quad\text{and}\quad
U_d = \left(\begin{array}{rr} c_d &  s_d \cr
                              -s_d & c_d \end{array}\right).
\end{equation}
One has $\tilde U_d = c_d,\quad m_n = m_s,\quad m_{n-1}=m_d,\quad
\underline{M_d} = \underline{D_d} = m_d$. \l % ;\l
 $\tilde A$ is defined by
$\widetilde{U_d D_d U_d^T} \equiv (U_d D_d U_d^T)_{11}=
 \tilde U_d \underline{D_d} \tilde U_d^T + m_s \tilde A$,
and one has  explicitly
$\widetilde{U_d D_d U_d^T} = c_d^2 m_d + s_d^2 m_s$,
$\tilde U_d \underline
{D_d} \tilde U_d^T = c_d^2 m_d$. % ; 
This yields $\tilde A = s_d^2$.
The condition (\ref{eq:gen1}) 
% writes
takes the form
\begin{equation}
m_d = c_d^2 m_d + s_d^2 m_s -\epsilon_d\;m_d
\quad \text{or}\quad
s_d^2 = \frac {\epsilon_d\;m_d}{m_s - m_d},\quad \text{with}
\quad\epsilon_d \leq 1,
\end{equation}
and, 
for $m_s \gg m_d$,
the small  $\sin \theta_d \approx \theta_d$
scales likes $\sqrt{\epsilon_d\,\frac{m_d}{m_s}}$ with $\epsilon_d
\leq 1$.
The same argument applied to $u$-type quarks yields:
$\theta_u$ scales like $\sqrt{\epsilon_u\,\frac{m_u}{m_c}}, \epsilon_u
\leq 1$ when $m_c \gg m_u$.

So, the (small) Cabibbo angle $\theta_c = \theta_d - \theta_u$ scales like
\begin{equation}
\theta_c \rightarrow \sqrt{\epsilon_d\,\frac{m_d}{m_s}}
 -\sqrt{\epsilon_u\,\frac{m_u}{m_c}},\quad \epsilon_d,\epsilon_u \leq 1,
 \quad\text{when}\quad m_s,m_c \gg m_d,m_u,
\label{eq:cabs}
\end{equation}
which is  well known \cite{Weinberg}  (see also \cite{Fritzsch} and references
therein) to be compatible with the experimental values of the quark masses.
This encouraging result we shall extend to the case of three
generations.             
%
%%%%%%%%%%%%%%%%%%%%%%%%%%%%%%%%%%%%%%%%%%%%%%%%%%%%%%%%%%%%%%%%%%%%%%%%%%%
\subsection{Example 2: the case of three generations}
\label{subsection:gen3gen}
One has
\begin{equation}
M_d  = U_d D_d U_d^\dagger \quad\text{with}\quad
D_d = \left(\begin{array}{ccc} m_d & 0 & 0 \cr
                               0  & m_s & 0 \cr
                               0  &  0  & m_b
\end{array}\right)\quad\text{and}\quad
\underline{D_d} = \left(\begin{array}{cc}
                                             m_d & 0 \cr
                                              0  & m_s 
                          \end{array}\right).
\end{equation}
The general condition (\ref{eq:COND}) writes
\begin{equation}
\underline{M_d} = \underline{U_d}\underline{D_d} \underline{U_d}^\dagger 
= \widetilde{U_d D_d U_d^\dagger} - \epsilon_d\, m_s {\mathbb I}_2,\quad
\vert\epsilon_d\vert \leq 1.
\end{equation}
According to (\ref{eq:expand}), the equation
\begin{equation}
\widetilde{U_d D_d U_d^\dagger} = \tilde U_d \underline{D_d} \tilde
U_d^\dagger
  + m_b \left(\begin{array}{cc}
  \vert(U_d)_{13}\vert^2 & (U_d)_{13}\ol{(U_d)_{23}} \cr
        \ol{(U_d)_{13}}(U_d)_{23} & \vert(U_d)_{23}\vert^2
        \end{array}\right)
\end{equation}
defines $\tilde A$ as the $2\times 2$ matrix factorizing $m_b$.
(\ref{eq:gen2}) writes now:
\begin{equation}
\tilde A \equiv \left( \begin{array}{cc}
    \vert(U_d)_{13}\vert^2 & (U_d)_{13}\ol{(U_d)_{23}}\cr
    \ol{(U_d)_{13}}(U_d)_{23} & \vert(U_d)_{23}\vert^2
        \end{array}\right)
= \frac
{ \underline{U_d} \underline{D_d} \underline{U_d}^\dagger
- \tilde{U_d}  \underline{D_d} \tilde{U_d}^\dagger + \epsilon_d\, m_s {\mathbb
  I}_2 } {m_b}, \quad \vert\epsilon_d\vert \leq 1.
\label{eq:bound31}
\end{equation}
All elements of the unitary matrices $U_d$ and $\underline{U_d}$ having
moduli bounded  by $1$, and the highest mass scale occurring in the
numerator of the l.h.s. of (\ref{eq:bound31}) being $m_s$, one gets
\begin{equation}
[\vert(U_d)_{13}\vert, \vert(U_d)_{23}\vert] \approx [\eta_{13}, \eta_{23}]
\sqrt{\frac{m_s}{m_b}},
\label{eq:Ud1}
\end{equation}
where $\eta_{13}, \eta_{23}$ are coefficients at most equal to $2(n-1)^2 + 1
\equiv 9$.
The unitarity of $U_d$ ensures that the same bounds occur for
$\vert(U_d)_{31}\vert, \vert(U_d)_{32}\vert$. 
The same argument applied to $u$-type quarks entails that
\begin{equation}
[\vert(U_u)_{13}\vert,\vert(U_u)_{23}\vert,\vert(U_u)_{31}\vert,\vert(U_u)_{32}]\vert \approx
[\zeta_{13},\zeta_{23},\zeta_{31},\zeta_{32}]\sqrt{\frac{m_c}{m_t}},
\label{eq:Uu1}
\end{equation}
with $\zeta$ coefficients at most equal to $9$.
Equations (\ref{eq:Ud1}) - 
(\ref{eq:K1}), being direct consequences
of the general requirement (\ref{eq:COND}),
have been  obtained independently of the existence of any mass hierarchy.
 However, it is clear that they can be of no use when there is no
 hierarchy.
When $m_s \ll m_b$, $m_c \ll m_t$, this
proves the hierarchical property of $U_u$ and $U_d$ in the general sense
defined in section \ref{section:defhier}. By the properties emphasized
there and the definition (\ref{eq:CKM}), it entails
that $K$ is hierarchical according to:
\begin{equation}
[\vert K_{13}\vert ,\vert K_{23}\vert ,\vert K_{31}\vert ,\vert K_{32}\vert ]
\approx
[\beta_{13},\beta_{23},\beta_{31},\beta_{32}] \sqrt{\frac{m_c}{m_t}} \pm
[\delta_{13},\delta_{23},\delta_{31},\delta_{32}] \sqrt{\frac{m_s}{m_b}},
\label{eq:K1}
\end{equation}
with $\beta,\delta$ coefficients  formally not exceeding $9$ and
typically much smaller than $9$.
%
%%%%%%%%%%%%%%%%%%%%%%%%%%%%%%%%%%%%%%%%%%%%%%%%%%%%%%%%%%%%%%%%%%%%%%%%%%%
\section{Strengthening the bounds}
\label{section:sronger}
%%%%%%%%%%%%%%%%%%%%%%%%%%%%%%%%%%%%%%%%%%%%%%%%%%%%%%%%%%%%%%%%%%%%%%%%%%%
%
By two different methods, we have obtained two different types of
scaling behaviors for the Cabibbo angle. 
Using as input in section \ref{section:2gen} the existence of
hierarchies and the experimentally observed smallness of this angle,
we deduced in (\ref{eq:limthetac}) a scaling behaviour like the inverse of the
heaviest ($s$ and $c$) quarks. 
Using the milder ``stability'' requirement (\ref{eq:COND})
for Yukawa matrices
in section \ref{section:3gen}, and no particular input concerning
either hierarchies of masses or smallness of mixing angles,
we have instead deduced in (\ref{eq:cabs}) a behaviour like
the inverse of the square root of the heaviest quark masses.
This last result is certainly not optimal since very little
information has been used. The combined action of large
cancellations in the numerator of the r.h.s. of (\ref{eq:bound31})
and of a possible dependence of $\epsilon$ on the quark masses and
on $m_{ij}$ (the sole condition on it is that it is bounded by $1$)
can concur to strengthen the results obtained  there.
It is consequently instructive to consider 
the possibility of strengthening
the bounds by incorporating our experimental 
knowledge on mass hierarchies and
on the smallness of the mixing angles.
We study explicitly  the case of three generations.
It proves convenient, for the problem of interest,  to use
a triangular basis for the mass matrices \cite{KuoWu}.

As stated in property {\em (iv)} of section
\ref{section:basic}, the change of basis from hermitian to triangular
can be achieved by independent
right-handed rotations on $u$- and $d$-type quarks; it does not modify
the mixing matrix $K$.
Also, when  hierarchies exist,
the non-diagonal elements generated in the lower triangle
when going from upper triangular to hermitian mass matrices differ very
little from the transposed conjugates of the ones of the starting triangular
matrix \cite{Desai}, respecting in particular the ``stability'' requirement
\footnote{
The advantage of triangular basis \cite{KuoWu}\cite{Desai}
is that the diagonal elements of the mass matrices, identical, then, to
their eigenvalues, are, when hierarchies exist, very
close to the quark masses, and that their entries can be expressed in
terms of the  elements of the CKM  mixing matrix and of the quark masses.}
.

The triangular matrices for $u$ and $d$-type quarks we call $T_u$ and
$T_d$. Their diagonalisation writes
\begin{equation}
T_d T_d^\dagger  = U_d D_d^2 U_d^\dagger ,
\quad T_u T_u^\dagger  = U_u D_u^2 U_u^\dagger .
\label{eq:diagT}
\end{equation}
Let us work, for example, in the $d$ sector, with
\begin{equation}
T_d = \left(\begin{array}{ccc}
             m_{11} & m_{12}\,e^{i\phi_{12}} & m_{13}\,e^{i\phi_{13}} \cr
                                0    & m_{22} & m_{23}\,e^{i\phi_{23}} \cr
                                0    &    0   & m_{33}
         \end{array}\right),\quad 
U_d = (U_d)_{ij},\ i,j= 1\cdots 3, \quad  D_d = diag(m_d,m_s,m_b),
\label{eq:tdud}
\end{equation}
where the $m_{ij}$ are real coefficients.
Equation (\ref{eq:diagT}) determines 
the unknown $(U_d)_{ij}, m_d,m_s,m_b$ as functions
of the independent parameters $m_{ij}, \phi_{ij}$.
In a Wolfenstein-like parameterization of $K$ \cite{Wolfenstein}
 and with $\lambda \approx .22$,
which includes the information that mixing angles are small,
(\ref{eq:diagT}), (\ref{eq:tdud}) yield 
respectively for
$(T_dT^\dagger_d)_{33}$, $(T_dT^\dagger_d)_{23}$ and $(T_dT^\dagger_d)_{13}$
(see \cite{KuoWu})
\begin{eqnarray}
&& m_{33}^2 = m_b^2 \vert(U_d)_{33}\vert^2
                    \left(1 + {\cal O}(\lambda^8)\right),\cr
&& m_{33}m_{23}\,e^{i\phi_{23}} = m_b^2 \ol{(U_d)_{33}}(U_d)_{23}
           \left(1 + {\cal O}(\lambda^4)\right),\cr  
&& m_{33}m_{13}\,e^{i\phi_{13}} = m_b^2 \ol{(U_d)_{33}} (U_d)_{13}
       \left(1 + {\cal O}(\lambda^4)\right),
       \label{eq:trieq}
\end{eqnarray}
and one finally gets
\begin{eqnarray}
m_b(m_{ij},\phi_{ij}) &\approx&
      \frac{\vert m_{33}\vert}{\vert(U_d)_{33}\vert(m_{ij},\phi_{ij})}
                   \left(1+ {\cal O}(\lambda^8)\right),\cr
\vert(U_d)_{23}(m_{ij},\phi_{ij})\vert &\approx&
      \frac{\vert m_{23}\vert}{m_b(m_{ij},\phi_{ij})}
                   \left(1+ {\cal O}(\lambda^4)\right),\cr
\vert(U_d)_{13}(m_{ij},\phi_{ij})\vert &\approx&
        \frac{\vert m_{13}\vert}{m_b(m_{ij},\phi_{ij})}
	           \left(1+ {\cal O}(\lambda^4)\right).
\label{eq:result}
\end{eqnarray}
The last two equations of (\ref{eq:result}) provide a $1/m_{b,t}$ behaviour
only if the mass $m_b$ in the denominator can be made very large independently
of the non-diagonal elements $m_{13}$ and $m_{23}$ of the Yukawa matrix
$T_d$.  We have seen in section \ref{section:2gen} that it is what occurs for
two generations.

The authors of the first reference in \cite{KuoWu} have shown that the
inversion of the system of equations in (\ref{eq:trieq}) plus those
associated with the other entries of $(T_dT_d^\dagger)$ yields,
in particular 
$m_b = m_{33}(1 + {\cal O}(\lambda^4)).$
Accordingly, the limit of very large $m_b$ depends only on
$m_{33}$, and not on $m_{13}$ and $m_{23}$.
The same argument can be made for the $u$-type quarks, and, (\ref{eq:CKM})
yields a result for the CKM mixing matrix $K$.

In the presence of hierarchies the $u$- and $d$- type quarks
($m_b$ and $m_t$ are much larger than the other quark masses)
and when the mixing angles are small, 
the non-diagonal ``external'' matrix elements
$\vert K_{13}\vert,\vert K_{23}\vert,\vert K_{31}\vert,\vert K_{32}\vert$
scale like
\begin{equation}
[\vert K_{13}\vert,\vert K_{23}\vert,\vert K_{31}\vert,\vert K_{32}\vert]
\approx
\frac{[\mu_{13},\mu_{23}, \mu_{31}, \mu_{32}]}{m_b} \pm
\frac{[\nu_{13},\nu_{23}, \nu_{31}, \nu_{32}]}{m_t}
\label{eq:resfin}
\end{equation}
where the $\mu$'s and $\nu$'s are independent mass scales.
%
%%%%%%%%%%%%%%%%%%%%%%%%%%%%%%%%%%%%%%%%%%%%%%%%%%%%%%%%%%%%%%%%%%%%%%%%%%%
\section{Conclusion}
\label{section:conclusion}
%%%%%%%%%%%%%%%%%%%%%%%%%%%%%%%%%%%%%%%%%%%%%%%%%%%%%%%%%%%%%%%%%%%%%%%%%%%
%
We have shown that, under a very
general assumption concerning the stability 
of Yukawa matrices as one steps
up or down generations, the non-diagonal border elements (lowest line and right
column) of the CKM mixing matrix scale like the inverse square root
of the largest masses of the upper generation of fermions.
For two generations, one recovers a well known 
behaviour for the Cabibbo angle.
A more constraining behaviour like
the inverse of the heaviest masses 
instead of their square root has been obtained by 
using the existence of mass hierarchies and the
smallness of the mixing angles.
%
%%%%%%%%%%%%%%%%%%%%%%%%%%%%%%%%%%%%%%%%%%%%%%%%%%%%%%%%%%%%%%%%%%%%%%%%%%%
\vskip 2mm
\begin{em}
\underline {Acknowledgements}. We are indebted to M.B. Gavela, with
whom this work initiated. It is a pleasure to thank B. Stech and
A. Romanino  for discussions and suggestions. 
S.T.P. acknowledges with gratefulness the hospitality of 
LPTHE, Universit\'e. de Paris VI, where part of the work on 
the present study was done. B.M. wants to
thank SISSA, where this work was completed, 
for the kind hospitality provided to him.
\end{em}
%
%%%%%%%%%%%%%%%%%%%%%%%%%%%%%%%%%%%%%%%%%%%%%%%%%%%%%%%%%%%%%%%%%%%%%%%%%%%%%
%%%%%%%%%%%%%%%%%%%%%%%%%%%%%%%%%%%%%%%%%%%%%%%%%%%%%%%%%%%%%%%%%%%%%%%%%%%%%
%
% \newpage\null
%
\begin{em}

\end{em}

\begin{thebibliography}{50}
%
\parskip = 0pt
\baselineskip = 13pt
%
\medskip
\bibitem{GlashowSalamWeinberg}
       S.L. GLASHOW: Nucl. Phys. 22 (1961) 579;\l
       A. SALAM: in ``Elementary Particle Theory: Relativistic Groups and
             Analyticity'' (Nobel symposium No 8), edited by N. Svartholm
             (Almquist and Wiksell, Stockholm 1968);\l
       S. WEINBERG:  Phys. Rev. Lett. 19 (1967) 1264.   
%
\bibitem{Pokorski}
M. OLECHOWSKI \& S. POKORSKI: Phys. Lett. B 257 (1991) 388.
%
\bibitem{RamondRobertsRoss}
P. RAMOND, R.G. ROBERTS \& G.G. ROSS: Nucl. Phys. B 406 (1993) 19.
%
\bibitem{PecceiWang}
R.D. PECCEI \& K. WANG: Phys. Rev. D 53 (1996) 2712.
%
\bibitem{GeorgiJarlskog}
H. GEORGI \& C. JARLSKOG: Phys. Lett. B 86 (1979) 297.
%
\bibitem{Fritzsch}
H. FRITZSCH:  Phys. Lett. B 70 (1977) 426,
B 184 (1987) 391,
and B 189 (1987) 191.
%
\bibitem{FritzschXing}
H. FRITZSCH \& Zhi-zong XING: Phys. Lett. B 413 (1997) 396.
%
\bibitem{FalconeTramontano}
D. FALCONE \& F. TRAMONTANO:  Phys. Rev. D59 (1999) 017302.
%
\bibitem{KuoWu}
T.K. KUO, S.W. MANSOUR \& G.H. WU: Phys. Rev. D 60: (1999) 093004,
and Phys. Lett. B 467 (1999) 116;
S.H. CHIU, T.K. KUO \& G.H. WU:  Phys. Rev. D 62 (2000) 053014.
%
\bibitem{Desai}
B.R. DESAI \& A. VAUCHER: 
  ``Quark Mass Matrices with Four and Five Texture
  Zeroes, and CKM Matrix, in terms of Mass Eigenvalues'', 
hep-ph/000723.
%
\bibitem{Branco}
G.C. BRANCO, D. EMMANUEL-COSTA \& R. GONZ\'ALEZ 
FELIPE: Phys. Lett. B477 (2000) 147.
%
\bibitem{Berger}
M.S. BERGER \& K. SIYEON: 
 ``Discrete Flavor Symmetries and Mass Matrix
 Textures'', 
hep-ph/0005249.
%
\bibitem{BarbieriHallRomanino}
R. BARBIERI, L.J. HALL \& A. ROMANINO:  Phys. Lett. B 401 (1997) 47.
%
\bibitem{CKM}
N. CABIBBO: Phys. Rev. Lett. 10 (1963) 531;\l
M. KOBAYASHI and T. MASKAWA:  Prog. Theor. Phys. 49 (1973) 652.
%
\bibitem{Wolfenstein}
L. WOLFENSTEIN: Phys. Rev. Lett. 51 (1983) 1945.
%
\bibitem{TurnbullAitken}
F.W. TURNBULL \& A.C. AITKEN: ``An Introduction to the Theory of Canonical
Matrices'' (Dover, New York, 1961) p. 194.
%
\bibitem{FramptonJarlskog}
P.H. FRAMPTON \& C. JARLSKOG:  Phys. Lett. 154 B (1985) 421.
%
\bibitem{FeruglioMasieroMaiani}
F. FERUGLIO, A. MASIERO \& L. MAIANI: Nucl. Phys. B 387 (1992) 523;\l
F. FERUGLIO: 
 ``The physics of the chiral fermions'', 
hep-ph/9405260.
%
\bibitem{Weinberg}
S. WEINBERG:  Trans. NY. Acad. Sci. (Ser. II) 38 (1977) 185.
%
\end{thebibliography}
\end{document}